\documentclass[11pt,twoside,showpacs, preprint]{revtex4}
\usepackage{graphicx,latexsym,amssymb,amsmath,color, multirow, mathrsfs, booktabs, helvet}

\newcommand{\beq}{\begin{equation}}
\newcommand{\eeq}{\end{equation}}
\newcommand{\bsub}{\begin{subequations}}
\newcommand{\esub}{\end{subequations}}
\newcommand{\cals}[1]{{\mathscr #1}}
\newcommand{\scr}[1]{{\mathscr #1}}
\newcommand{\lrb}[1]{\left(#1\right)}

\newcommand{\lrs}[1]{\left[#1\right]}

\newcommand{\svec}[1]{{\mbox{\boldmath${ #1}$}}}
\newcommand{\ivec}{\vec}
\newcommand{\ff}[1]{\frac{1}{#1}}
\newcommand{\tabref}[1]{{\sf\bfseries Table \ref{#1}}}
\newcommand{\figref}[1]{{\sf\bfseries Fig. \ref{#1}}}
\newcommand{\mev}{\text{MeV}}
\newcommand{\brule}{\begin{ruledtabular}}
\newcommand{\erule}{\end{ruledtabular}}
\newcommand{\er}{{\text{R}}}
\newcommand{\nr}{{\text{NR}}}

\begin{document}
\title{Density-Dependent Relativistic Hartree-Fock Approach }
\author{WenHui Long}
 \affiliation{School of Physics, Peking University, Beijing 100871,
China}
 \affiliation{Institut de Physique Nucl$\acute{e}$aire, CNRS-IN2P3, Universit$\acute{e}$
Paris-Sud, 91406 Orsay, France}
\author{Nguyen Van Giai}
 \affiliation{Institut de Physique Nucl$\acute{e}$aire, CNRS-IN2P3, Universit$\acute{e}$ Paris-Sud,
91406 Orsay, France}
\author{Jie Meng }
 \affiliation{School of Physics, Peking University, Beijing 100871, China}
 \affiliation{Institute of Theoretical Physics, Chinese Academy of Sciences, Beijing 100080, China}
 \affiliation{Center of Theoretical Nuclear Physics, National Laboratory of Heavy Ion Accelerator, Lanzhou 730000, China}

\begin{abstract}

A new relativistic Hartree-Fock approach with density-dependent
$\sigma$, $\omega$, $\rho$ and $\pi$ meson-nucleon couplings for
finite nuclei and nuclear matter is presented. Good description for
finite nuclei and nuclear matter is achieved  with a number of
adjustable parameters comparable to that of the relativistic mean
field approach.  With the Fock terms, the contribution of the
$\pi$-meson is included and the description for the nucleon
effective mass and its isospin and energy dependence is improved.

\end{abstract}

 \maketitle

The relativistic mean field (RMF) theory \cite{Miller:1972, Walecka:1974} has received much
attention due to its successful description of numerous nuclear phenomena \cite{Serot:1986,
Reinhard:1989, Ring:1996, Serot:1997, Bender:2003, Vretenar:2005, Meng:2005}. In its most widely
employed versions, i.e., with either self-coupling interactions or density-dependent meson-nucleon
couplings, the RMF theory with a limited number of parameters can describe very well a very large
amount of data: saturation properties of nuclear matter \cite{Brockmann:1990}, nuclear binding
energies and radii, the isotopic shifts in the Pb-region \cite{Sharma:1993b}. It gives a natural
description of the nuclear spin-orbit potential \cite{Lalazissis:1998}, and explains the origin of
the pseudospin symmetry \cite{Arima:1969, Hecht:1969} and spin symmetry of the anti-nucleon
spectrum \cite{Zhou03prl} as a relativistic symmetry \cite{Ginocchio97,Meng98r,Meng99prc,
Zhou03prl}. In spite of these success, there are still a number of questions needed to be answered
in the RMF theory: the contributions due to the exchange (Fock)
terms and the pseudo-scalar $\pi$-meson. 

There exist attempts to include the exchange terms in the
relativistic description of nuclear matter and finite nuclei. The
earlier  relativistic Hartree-Fock (RHF) method led to underbound
nuclei due to the missing of the meson self-interactions
\cite{Bouyssy:1987}. Further developments were made by taking into
account approximately the nonlinear self-couplings of the
$\sigma$-field \cite{Bernardos:1993} or by introducing the products
of six and eight nucleon spinors in the zero-range limit
\cite{Marcos:2004}. Although some improvements were obtained, the
RHF method is still not comparable with the RMF theory in the
quantitative description of nuclear systems. The relativistic point
coupling model has been used to investigate nuclei systems
\cite{Burvenich:2002} and the consequences of Fierz transformations
acting upon the contact interactions for nucleon fields occurring in
relativistic point coupling models has been investigated in Hartree
approximation, which yield the same models but in Hartree-Fock
approximation instead \cite{Sulaksono:2003, Madland:2004}. It has
been suggested that the Hartree-Fock approximation may constitute a
physically more realistic framework for power counting and QCD
scaling than the Hartree approximation.

In this work, a new RHF approach which contains density-dependent
meson-nucleon couplings is developed. With a number of adjustable
parameters comparable to that of RMF Lagrangians, this
density-dependent RHF (DDRHF) theory can give a good description of
nuclear systems without dropping the Fock terms. Furthermore,
important features like the behavior of neutron and proton effective
masses \cite{Jaminon:1989} can be interpreted well in DDRHF in
comparison with the results of non-relativistic
Brueckner-Hartree-Fock (BHF) \cite{Zuo:2005} and
Dirac-Brueckner-Hartree-Fock (DBHF) calculations \cite{Ma:2004,
Dalen:2005}.

The most important parts of the nuclear force are the short-range
repulsive and medium-range attractive components. In analogy with
the strong interaction in free space which is described by meson
exchanges, it is convenient to represent the strong interaction in
nuclear medium by the exchange of effective isoscalar and isovector
mesons. The description of nucleon and meson degrees of freedom has
to rely ultimately on the relativistic quantum field approach.
According to this spirit, we start from an effective Lagrangian
density $\cals L$ constructed with the degrees of freedom associated
with the nucleon field($\psi$), two isoscalar meson fields ($\sigma$
and $\omega$), two isovector meson fields ($\pi$ and $\rho$) and the
photon field ($A$). The parameters of the model are the effective
meson masses and meson-nucleon couplings.

With the general Legendre transformation
 \beq
\cals H = T^{00} = \frac{\partial\cals L}{\partial \dot\phi_i}\dot\phi_i - \cals L,
 \eeq
one can obtain the effective Hamiltonian from the Lagrangian density $\cals L$ as
 \beq\label{Hamiltonian}\begin{split}
\scr H = &\bar\psi\lrb{-i\svec\gamma\cdot\svec\nabla + M}\psi \\
+\ff2 &\int d^4 x_2 \sum_{{i=\sigma,\omega},\atop\rho,\pi,
A}\bar\psi(x_1) \bar\psi(x_2) \Gamma_i D_i(x_1,x_2)
\psi(x_2)\psi(x_1),
 \end{split}\eeq
where $D_i(x_1,x_2)$ represent the corresponding meson propagators,
and the interaction vertices $\Gamma_i$ are defined as,
 \bsub\begin{align}
\Gamma_\sigma(1,2)\equiv& -g_\sigma(1) g_\sigma(2),\\
\Gamma_\omega(1,2)\equiv&+g_\omega(1)\gamma_\mu(1) g_\omega(2)\gamma^\mu(2),\\
\Gamma_\rho(1,2)\equiv& +g_\rho(1)\gamma_\mu(1)\ivec\tau(1)\cdot
g_\rho(2)\gamma^\mu(2)\ivec\tau(2),\\
\Gamma_\pi(1,2)\equiv& -\lrs{\textstyle{\frac{f_\pi}{m_\pi}}\ivec\tau\gamma_5\gamma_\mu\partial^\mu
}_1\cdot\lrs{\textstyle{\frac{f_\pi}{m_\pi}}\ivec\tau\gamma_5\gamma_\nu\partial^\nu }_2,\\
\Gamma_A(1,2)\equiv&+ \frac{e^2}{4}\lrs{\gamma_\mu(1-\tau_3)}_1 \lrs {\gamma^\mu(1-\tau_3)}_2.
 \end{align}\esub

Following the experience and success in DDRMF \cite{Brockmann:1992, Lenske:1995, Fuchs:1995,
Typel:1999, Niksic:2002, Long04, Vretenar:2005}, the meson-nucleon couplings $g_\sigma, g_\omega,
g_\rho$ and $f_\pi$ are taken as functions of the baryonic
density $\rho_b$.
For $\sigma$- and $\omega$-meson, the density-dependence of the
couplings $g_\sigma$ and $g_\omega$ are chosen as
 \beq
g_i(\rho_b) =g_i(\rho_{0}) f_i(\xi),
 \eeq
where $i =\sigma, \omega$, and
 \beq
f_i(\xi ) = a_i\frac{1 + b_i(\xi +d_i)^2}{1+ c_i(\xi +d_i)^2},
 \eeq
is a function of $\xi =\rho_b/\rho_{0}$, and $\rho_{0}$ denotes the
baryonic saturation density of nuclear matter. In addition, five
constraint conditions $f_i(1) = 1, f_\sigma''(1) = f_\omega''(1)$,
and $f_i''(0) =0$ are introduced to reduce the number of free
parameters.
For simplicity, the exponential density-dependence is adopted for
$f_\pi$ as well as $g_\rho$~\cite{Jong:1998}:
 \bsub\begin{align}
g_\rho(\rho_b)  =& g_\rho(0) e^{-a_\rho\xi}, \\
f_\pi(\rho_b)   =& f_\pi(0) e^{-a_\pi\xi}.
 \end{align}\esub
 The coupling constants
$g_\rho(0)$ and $f_\pi(0)$ are fixed to their values in free space.
One reason to fix the coupling constants $g_\rho(0)$ and $f_\pi(0)$
to their values in free space is just to reduce the number of free
parameters and another reason is that the inclusion of Fock terms
allows such choice. There are in total 8 free parameters, i.e.,
$m_\sigma$, $g_\sigma (\rho_{0})$, $g_\omega (\rho_{0})$, $a_\rho$,
$a_\pi$, and three others from the density-dependence of $g_\sigma$
and $g_\omega$. A new parametrization called PKO1 is found (see
\tabref{tab:PKO1}) by fitting the masses of the nuclei $^{16}$O,
$^{40}$Ca, $^{48}$Ca, $^{56}$Ni, $^{68}$Ni, $^{90}$Zr, $^{116}$Sn,
$^{132}$Sn, $^{182}$Pb, $^{194}$Pb, $^{208}$Pb and $^{214}$Pb, and
the values of the baryonic saturation density $\rho_0$, the
compression modulus $K$ and the symmetry energy $J$ of nuclear
matter at the saturation point.

It should be emphasized that here the effective interaction PKO1 is
obtained by fitting the empirical properties of nuclei and the
nuclear matter at the saturation point.  In Table I one finds
$a_\rho$ = 0.076 and $a_\pi$  = 1.232. This means that $g_\rho(1)/
g_\rho (0)$ =0.93 and $f_\pi (1)/ f_\pi (0)$ =0.36, i.e., the
contribution from the pion is reduced heavily than that in free
space. In fact, the effect of pion has been taken into account
effectively via the other mesons. It will be refined in the future
if more information is used to constrain the density dependence of
the effective interaction in the medium.

 \begin{table}[htbp]\centering\setlength{\tabcolsep}{1em}
\caption{The effective interaction PKO1 for DDRHF with $M = 938.9$MeV, $m_\omega = 783.0\mev,
m_\rho = 769.0$MeV , $m_\pi = 138.0$MeV. }\label{tab:PKO1}
 \brule\begin{tabular}{c|r||c|r||c|r}
 $m_\sigma$& 525.7691& $a_\sigma$& 1.3845& $a_\omega$&1.4033 \\
$g_\sigma$&8.8332&$b_\sigma$&1.5132&$b_\omega$&2.0087\\
$g_\omega$&10.7299&$c_\sigma$&2.2966&$c_\omega$&3.0467\\
$g_\rho(0)$&2.6290&$d_\sigma$&0.3810&$d_\omega$&0.3308\\
$f_\pi(0)$&1.0000&$a_\rho$&0.0768&$a_\pi$&1.2320
 \end{tabular}\erule
 \end{table}

The PKO1 parameter set gives the following nuclear matter bulk properties: compression modulus
$K=250.24$MeV, symmetry energy $J=34.37$MeV, binding energy per particle $E/A=-15.996$MeV,
saturation baryonic density $\rho_0=0.1520$fm$^{-3}$.

For finite nuclei, the self-consistent Dirac equations are solved in
coordinate space with techniques similar to those used in RMF
 \cite{Meng:1996,Meng:1998a}. The non-local exchange (Fock) potentials are treated
exactly as in Ref. \cite{Bouyssy:1987}. Calculations are carried out
for a set of selected nuclei (S.N.), i.e., $^{16}$O, $^{40}$Ca,
$^{48}$Ca, $^{56}$Ni, $^{58}$Ni, $^{68}$Ni, $^{90}$Zr, $^{112}$Sn,
$^{116}$Sn, $^{124}$Sn, $^{132}$Sn, $^{182}$Pb, $^{194}$Pb,
$^{204}$Pb, $^{208}$Pb, $^{214}$Pb, and $^{210}$Po, as well as the
Sn and Pb isotopic chains. For the open shell nuclei, the pairing
correlations are treated by the BCS method with a density-dependent
delta force \cite{Dobaczewski:1996}. A detailed comparison with the
predictions of some typical RMF parameterizations: PK1
\cite{Long04}, PKDD \cite{Long04}, NL3 \cite{Lalazissis:1997} and
DD-ME1 \cite{Niksic:2002} are summarized in \tabref{tab:rmsd} where
the root mean square (rms) deviations from the data are shown.  As
one can see in \tabref{tab:rmsd}, the DDRHF approach with PKO1
provides a good quantitative description of finite nuclei, sometimes
better than the RMF approach.  It should be emphasized that this is
the first time for the RHF approach to provide such a good
quantitative description for the finite nuclei and nuclear matter.

\begin{table}[htbp]
\caption{The rms deviations $\Delta$  from the data for the RHF
calculations with PKO1 in comparison with that of RMF with PK1,
PKDD, NL3 and DD-ME1. The rows from two to ten are respectively: the
binding energies $E_b$ of the selected nuclei ({S.N.}) and the
even-even nuclei in {Pb} and {Sn} chains; two-neutron separation
energies $S_{2n}$ of Pb and Sn isotopes; charge radii $r_c$ of the
S.N. and Pb isotopes; isotope shifts ({I.S.}) of Pb isotopes;
spin-orbit ({S.O.}) splittings of doubly magic
nuclei.}\label{tab:rmsd}
 \brule
 \begin{tabular}{c|c|ccccc}
&&{PKO1}&PK1&PKDD&NL3&DD-ME1\\ \hline
\multirow{3}{0.8cm}{$\Delta_{E_b}$}&S.N.& {1.6177}&1.8825&2.3620&2.2506&2.7561\\
&Pb&  {1.8995}&2.0336&2.7007&2.0021&2.1491\\
&Sn&  {1.2665}&1.9552&2.4567&1.6551&0.9168\\
\hline \multirow{2}{0.8cm}{$\Delta_{S_{2n}}$}& Pb&  {0.6831}&0.9192&1.3139&0.9359&1.2191
\\
&Sn& {0.6813}&0.7762&1.0629&0.8463&0.7646 \\ \hline
\multirow{2}{0.8cm}{$\Delta_{r_c}$}&S.N.& {0.0269}&0.0204&0.0188&0.0177&0.0163\\
&Pb& {0.0056}&0.0061&0.0060&0.0143&0.0150\\ \hline $\Delta_{\text{I.S.}}$&Pb&
{0.0760}&0.0784&0.0784&0.0679&0.0567\\\hline
\multirow{5}{0.8cm}{$\Delta_{\text{S.O.}}$}&
  O& 0.1761&0.2879&0.6817&0.2195&0.1107\\
&Ca&0.5078&0.6638&0.8159&0.7184&0.6041\\
&Ni&0.3959&0.9923&1.3287&1.3315&0.9029\\
&Sn&0.1650&0.3300&0.6913&0.4757&0.5408\\
&Pb&0.2014&0.3902&0.6370&0.4604&0.4588\\
\end{tabular}\erule
\end{table}

\begin{figure}[htbp]
\includegraphics[width=7.0cm]{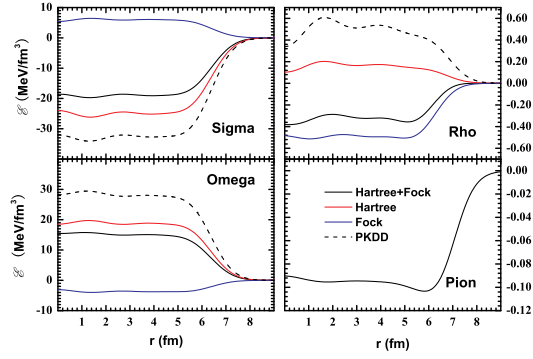}
\caption{Energy density contributions from Hartree and Fock terms in
different channels for neutrons in $^{208}$Pb given by DDRHF with
PKO1, in comparison with RMF with PKDD .} \label{fig:PRLEnD4}

\end{figure}

From previous discussion, one can find that good description of
nuclear systems comparable to that of RMF can be obtained without
dropping the Fock terms. Taking $^{208}$Pb as an example, shown in
\figref{fig:PRLEnD4} are the neutron energy densities from Hartree
and Fock terms in different meson channels in DDRHF, compared with
the results of RMF with PKDD \cite{Long04}. There exist significant
and remarkable differences between  DDRHF and RMF results.

Although the attractive and repulsive parts of the nuclear force are
mainly provided by $\sigma$- and $\omega$-mesons respectively, the
contributions in DDRHF are much less than their corresponding ones
in RMF, as shown in \figref{fig:PRLEnD4}. For the isovector
channels, the isovector $\rho$- and $\pi$-mesons in the DDRHF
approach become attractive due to the stronge Fock terms. While in
the standard RMF with $\sigma$-, $\omega$- and $\rho$-mesons, the
isospin part of nuclear force is provided only by the direct part of
$\rho$-meson, which gives the repulsive interaction for the
neutrons. Furthermore it should be emphasized that one of the
advantage of the DDRHF is the inclusion of the $\pi$-meson which is
very important at large distance in DDRHF.

An important difference between the RHF and RMF approaches is the
nucleon effective mass. In the medium, particles or quasi-particles
behave as if their mass is different from their bare mass due to
interactions with surrounding particles, which is reflected in the
level density as an example. In Ref. \cite{Jaminon:1989}, the
nucleon effective mass has been discussed and it is shown that there
are two sources of modification of the bare mass:  the non-locality
of the mean field which gives rise to the so-called k-mass $M^*_k$,
and the energy dependence of the mean field which leads to the
E-mass $M^*_E$. The total effective mass $M^*$ is related to $M^*_k$
and $M^*_E$. One can already note that the RMF (RHF) mean field is
local (non-local) in coordinate space and therefore, it can be
expected that their effective masses will differ. It should be also
emphisized that, in RMF theory appears the Lorentz scalar mass $M_S
= M+\Sigma_S$ where $\Sigma_S$ is the scalar self-energy. It should
not be confused with any of the $M^*$ and one should refer to it as
the scalar mass, or Dirac mass.

\begin{figure}[htbp]
\includegraphics[width=7.0cm]{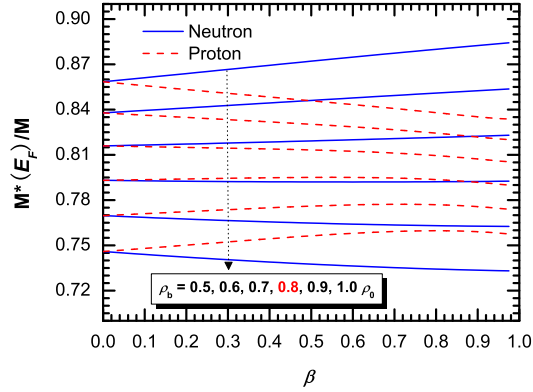}
\caption{ Neutron and proton effective masses $M_\nr^*$ at their
corresponding Fermi energy $E_F$ calculated in DDRHF with PKO1 as
functions of $\beta=(N-Z)/A$ for different baryonic densities.}
\label{fig:NRISFDa}
\end{figure}

\begin{figure}[htbp]
\includegraphics[width=7.0cm]{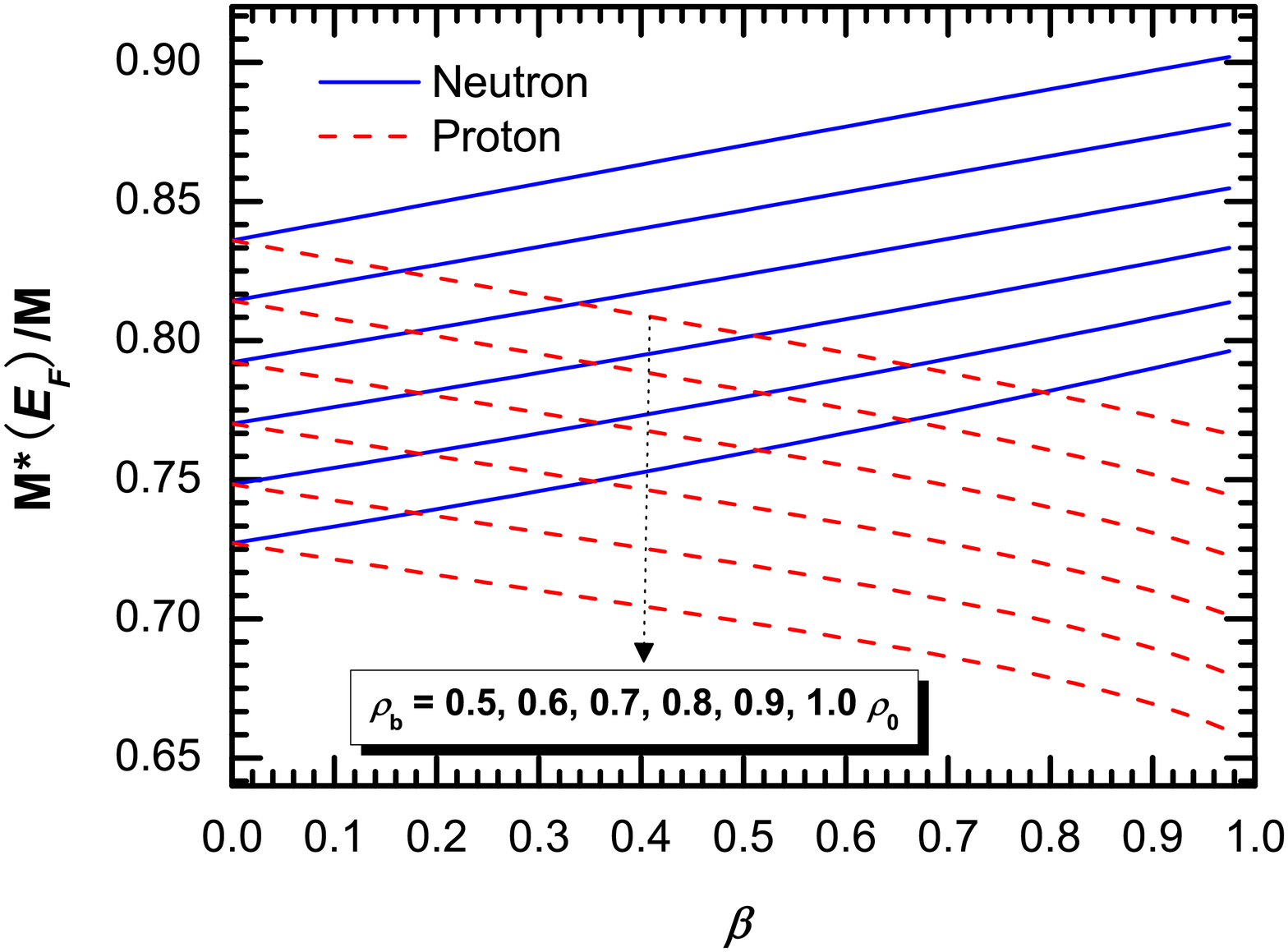}
\caption{ Same as Fig.2, but for $M_\er^*$. } \label{fig:RISFDa}
\end{figure}

In the non-relativistic framework, the energy-momentum relation
 \beq\label{Non-FM}
\frac{1}{2M} k^2 +  V(k;\epsilon) = \epsilon
 \eeq
leads to the effective mass $M^*$ \cite{Jaminon:1989}:
 \begin{eqnarray}
 \frac{M^*}{M} &\equiv& 1 - \frac{d
 V(k(\epsilon);\epsilon)}{d\epsilon},
 \end{eqnarray}
where $\epsilon = E - M$ is the particle energy and $ V(k;\epsilon)$
is the momentum- and energy-dependent mean field.

In a relativistic framework like RMF or RHF, the energy-momentum
relation is,
 \beq\label{kmeStar}
({\svec k}+\svec{\hat k}\Sigma_V)^{2} + (M+\Sigma_S)^{2} = (E-\Sigma_0)^{2},
 \eeq
where $\Sigma_S$, $\Sigma_V$, and $\Sigma_0$ are respectively the
scalar, spacelike- and timelike-vector components of the
self-energy. Its Schr\"odinger-type form can be derived as:
 \beq\label{Re-FM}
\frac{1}{2M}k^2 +  V(k;\epsilon) - \frac{\epsilon^2}{2M} = \epsilon,
 \eeq
which give the effective masses ${M_\er^*}$,
 \beq\label{Rmass}
  \frac{M_\er^*}{M} = 1-\frac{d}{d\epsilon} \lrs{ V(k(\epsilon);\epsilon) -
  \frac{\epsilon^2}{2M}}
 \eeq
and $M_\nr^* = M_\er^* - \epsilon$ in the non-relativistic
approximation by neglecting the last term at the left side of
Eq.(\ref{Re-FM}). One can see that $M^*_{NR}$ is the effective mass
in Refs. \cite{Jaminon:1989, Ma:2004} and $M^*_R$ the group mass in
Ref. \cite{Jaminon:1989}, and they are the same in the
non-relativistic approach. In the relativistic approach they can be
significantly different, as shown in the following.

The neutron and proton effective masses $M^*_{NR}$ and $M^*_{R}$ at
their corresponding Fermi energy $E_F$ from the DDRHF calculations
with PKO1 are respectively shown as functions of  $\beta=(N-Z)/A$ in
\figref{fig:NRISFDa} and \figref{fig:RISFDa} for different density
$\rho_b$. At lower density, the RHF gives the trend that
$M_{\nr,n}^*(E_{F,n})
> M_{\nr,p}^*(E_{F,p})$
but this trend is  reversed around $0.8\rho_0$. For $M^*_R$, one
always have  $M_{\er,n}^*(E_{F,n})
> M_{\er,p}^*(E_{F,p})$ for all densities.

\begin{figure}[htbp]
\includegraphics[width=7.0cm]{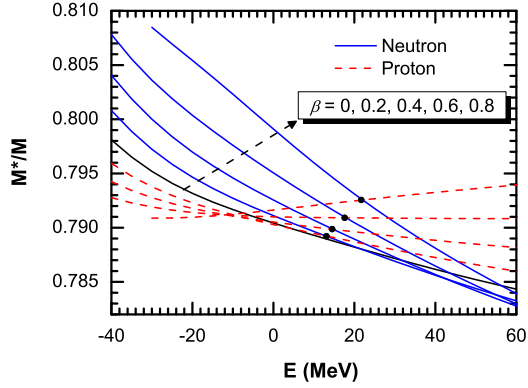}
\caption{The energy dependence of the effective mass $M^*_{NR}$
calculated in RHF with PKO1, for $\rho=0.8\rho_0$ and different
values of neutron excess $\beta$. } \label{fig:MNREISa}
\end{figure}

In contrast, for RMF, the relatively simple expressions tell us that
one always have $M^*_{\nr, n}<M^*_{\nr, p}$ and
$M_{\er,n}^*(E_{F,n}) > M_{\er,p}^*(E_{F,p})$ for neutron rich
system. This difference between RHF and RMF is related to the
presence of exchange (Fock) terms which bring non-locality effects
to the RHF self-energies. It is worthwhile to mention that in
Brueckner-Hartree-Fock studies it is found that $M^*_n (E_{F,n})
>M^*_p (E_{F,p})$ \cite{Bombaci:1991, Dalen:2005}, but at larger
density ($\rho_b = 0.17$fm$^{-3}$) \cite{Bombaci:1991}.

Another significant difference between RMF and DDRHF predictions is
the energy dependence of $M_\nr^*$. In RMF, $M_\nr^*$ is a constant
whereas it is a function of $E$ or $k$  in RHF. In
\figref{fig:MNREISa}, the energy dependence of $M_\nr^*$ for
different $\beta$ at $\rho_b = 0.8\rho_0$ is shown. The neutron
effective mass tends to decrease with the energy whereas the proton
mass is more constant or slightly increases in neutron rich matter.

One can see that the neutron effective masses are larger than the
proton ones at low energy, i.e., $M_{\nr,n}^*(E)>M_{\nr,p}^*(E)$,
and depending on the $\beta$, a different feature appears at energy
$E \sim 15-20~$MeV ( Solid points ). It was also found in Dirac
Brueckner-Hartree-Fock calculations that
$M_{\nr,n}^*(E)>M_{\nr,p}^*(E)$ \cite{Ma:2004}, but at high energy
($E = 50$MeV). Combining with the discussion of
\figref{fig:NRISFDa}, one can conclude that the DDRHF will predict
$M_{\nr,n}^* > M_{\nr,p}^*$ at low energy or low density while
$M_{\nr,n}^* < M_{\nr,p}^*$ for the RMF.

In summary, it has been demonstrated that one can go beyond the
standard relativistic mean field approach to include the exchange
(Fock) terms and the new couplings such as pion-nucleon couplings
which are effective only through exchange terms. These exchange
terms are the cause of subtle effects such as the isospin dependence
of the effective masses. The same (or even better) quantitative
description of nuclear properties comparable to RMF can be achieved
with a number of adjusted parameters. It will open the door to the
future investigation of nuclei by the relativistic
Hartree-Fock-Bogoliubov approach and the relativistic RPA on top of
RHF approximation.

\begin{acknowledgments}
This work is  partly supported by the National Natural Science
Foundation of China under Grant No. 10435010, and 10221003, the
Doctoral Program Foundation from the Ministry of Education in China.
\end{acknowledgments}


\end{document}